\documentclass{aa}  
\usepackage{url}
\usepackage{natbib,twoopt}
\usepackage{amsmath}

\usepackage{color}

\usepackage[colorlinks, linkcolor=blue, citecolor=blue, urlcolor=blue]{hyperref} 
\makeatletter
  \newcommandtwoopt{\intheteads}[3][][]{\href{http://adsabs.harvard.edu/abs/#3}%
    {\def\hyper@linkstart##1##2{}%
     \let\hyper@linkend\@empty\citealp[#1][#2]{#3}}}
  \newcommandtwoopt{\citepads}[3][][]{\href{http://adsabs.harvard.edu/abs/#3}%
    {\def\hyper@linkstart##1##2{}%
     \let\hyper@linkend\@empty\citep[#1][#2]{#3}}}
  \newcommandtwoopt{\citetads}[3][][]{\href{http://adsabs.harvard.edu/abs/#3}%
    {\def\hyper@linkstart##1##2{}%
     \let\hyper@linkend\@empty\citet[#1][#2]{#3}}}
  \newcommandtwoopt{\citeyearads}[3][][]%
    {\href{http://adsabs.harvard.edu/abs/#3}
    {\def\hyper@linkstart##1##2{}%
     \let\hyper@linkend\@empty\citeyear[#1][#2]{#3}}}
\makeatother

\usepackage{graphicx}
\usepackage{txfonts}
\begin{document}

   \title{Starburst-Driven Galactic Outflows}

   \subtitle{Unveiling the Suppressive Role of Cosmic Ray Halos}

   \author{Leonard E. C. Romano\inst{1,2,3}
          \and
          Ellis R. Owen\inst{4,5}
          \and
          Kentaro Nagamine\inst{4,6,7,8,9}
          }

   \institute{Universitäts-Sternwarte, Fakultät für Physik, Ludwig-Maximilians-Universität München, Scheinerstr. 1, D-81679 München, Germany \email{lromano@usm.lmu.de}
         \and
             Max-Planck-Institut für extraterrestrische Physik, Giessenbachstr. 1, D-85741 Garching, Germany
        \and
             Excellence Cluster ORIGINS, Boltzmannstr. 2, D-85748 Garching, Germany
        \and 
            Theoretical Astrophysics, Department of Earth and Space Science, The University of Osaka, 1-1 Machikaneyama, Toyonaka, Osaka 560-0043, Japan
         \and
            Astrophysical Big Bang Laboratory (ABBL), RIKEN Pioneering Research Institute (PRI), Wak\={o}, Saitama 351-0198, Japan
            \email{ellis.owen@riken.jp}
        \and
             Theoretical Joint Research, Forefront Research Center, The University of Osaka, 1-1 Machikaneyama, Toyonaka, Osaka 560-0043, Japan
        \and
            Kavli IPMU (WPI), UTIAS, The University of Tokyo, Kashiwa, Chiba 277-8583, Japan
        \and
            Department of Physics and Astronomy, University of Nevada, Las Vegas, 4505 S. Maryland Pkwy, Las Vegas, NV 89154-4002, USA
        \and
            Nevada Center for Astrophysics, 
            University of Nevada, Las Vegas, 4505 S. Maryland Pkwy, Las Vegas, NV 89154-4002, USA
             }

   \date{Received XXX; accepted YYY}

\abstract{}{
We investigate the role of cosmic ray (CR) halos in shaping the physical properties of starburst-driven galactic outflows.
}{
We construct a model for galactic outflows driven by a continuous central injection of energy, gas and CRs, where the treatment of CRs accounts for the effect of CR pressure gradients on the flow dynamics. 
The model parameters are set by the effective properties of a starburst. 
By analyzing the asymptotic behavior of our model, we derive the launching criteria for starburst-driven galactic outflows and determine their corresponding outflow velocities. 
}{
We find that, in the absence of CRs, stellar feedback can only launch galactic outflows if the star formation rate (SFR) surface density exceeds a critical threshold proportional to the dynamical equilibrium pressure. In contrast, CRs can always drive slow outflows. 
CR-driven outflows dominate in systems with SFR surface densities below the critical threshold, but their influence diminishes in highly star-forming systems.
However, in older systems with established CR halos, the CR contribution to outflows weakens once the outflow reaches the galactic scale height, making CRs ineffective in sustaining outflows in such environments.
}{
Over cosmic time, galaxies accumulate relic CRs in their halos, providing additional non-thermal pressure support that suppresses low-velocity CR-driven outflows. 
We predict that such low-velocity outflows are expected only in young systems that have yet to build significant CR halos. 
In contrast, fast outflows in starburst galaxies, where the SFR surface density exceeds the critical threshold, are primarily driven by thermal energy and remain largely unaffected by CR halos.
}{\vspace{3cm}} 

   \keywords{Galaxy winds -- circumgalactic medium -- cosmic rays -- starburst galaxies}

   \maketitle
%

\section{Introduction}
\label{sec:introduction} 

Galaxies with high star-formation surface densities often host large-scale outflow winds. Such winds have been observed in local starbursts, such as Arp 220, M82, and NGC 253 \citep[e.g.,][]{2013Natur.499..450B, 2015ApJ...814...83L, 2017ApJ...835..265W, 2018ApJ...853L..28B} and are widespread 
at high-redshifts, where galaxies are typically more compact and have higher star-formation rates relative to their stellar mass~\citep[see, e.g.][]{2019ApJ...886...29S, 2024ApJ...963...19N, 2024ARA&A..62..529T}. Outflow winds play an important role in redistributing energy, momentum, and baryons between the interstellar medium (ISM) and halos of galaxies. This makes them a key feedback component that regulates the evolution of galaxy ecosystems. Yet, despite their importance, a complete picture of the role they play remains unsettled \citep[see][for reviews]{2018Galax...6..114Z, 2024ARA&A..62..529T}. 

\begin{figure*}
\begin{center}
 \includegraphics[width=\linewidth, clip=true]{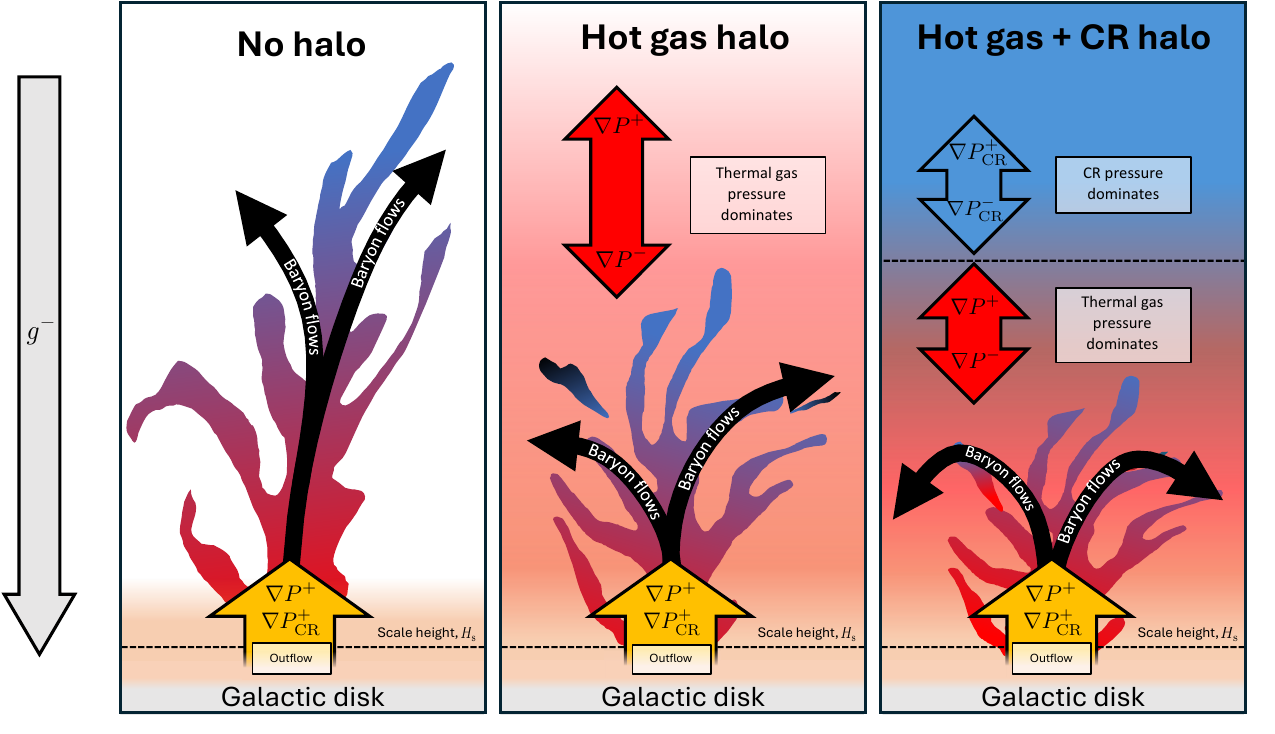}
 \end{center}
 \caption{Schematic of the structure of a starburst-driven outflow embedded in a galactic halo, propelled by thermal gas pressure and/or non-thermal CR pressure. The scale height of the warm ISM is indicated. All galaxies are embedded in a gravitational potential, $g$, which opposes the outflow. Superscripts + and - on quantities indicate whether each term contributes to or opposes the outflow, respectively. \textbf{Left}: In the absence of a substantial gas halo, outflowing gas is unconfined, allowing it to escape beyond the galaxy ecosystem against the galaxy’s gravitational potential. \textbf{Center}: As the galaxy builds up its stellar mass, feedback processes form a hot gas halo that suppresses the outflow, promoting baryonic recycling and enriching the CGM~\citep[see][]{2005ApJ...634L..37F, 2021ApJ...917...12S}. \textbf{Right}: When CRs are supplied to the halo, they accumulate over time, introducing a non-thermal halo component. Since non-thermal CR pressure gradients operate over larger length scales than thermal pressure gradients, an outflow erupting into the galaxy halo encounters distinct layers where thermally dominated and CR-dominated pressure gradients hinder its development. } 
 \label{fig:schematic_flows}
\end{figure*}

Detailed multi-wavelength observations of nearby starburst galaxies with outflows have revealed certain common features, including a bi-conical shape aligned along the minor axis of their host galaxy~\citep{2005ARA&A..43..769V}, extensions reaching 10s of kpc into the halo~\citep{2005ARA&A..43..769V, 2018Galax...6..114Z}, a terminal ``cap'' at a few kpc, e.g., at $\sim12$ kpc in M82 (see \citealt{1999ApJ...523..575L, 2007PASJ...59S.269T}), and the presence of entrained magnetic fields~\cite[e.g.][]{2019ApJ...870L...9J, 2021ApJ...914...24L} and high-energy CR particles~\citep[for an overview, see][]{2024Galax..12...22I}. 

Despite these apparent similarities, the physical configuration of individual galactic outflows can vary substantially. For instance, flow velocities ranging up to $\sim$1,000 km s$^{-1}$ have been reported
~\citep[e.g.][]{2013MNRAS.433..194B, 2015ApJ...809..147H, 2016A&A...588A..41C, 2016A&A...590A.125C, 2022ApJ...933..222X, 2024MNRAS.535.1684T}, while  densities and mass-loading factors span over 1.5 orders of magnitude~\citep[e.g.][]{2023ApJ...948...28X, 2015ApJ...809..147H}. 
This diversity can be attributed to differences in the energy, matter, and momentum being supplied to an outflow by its host galaxy \citep{2018Galax...6..114Z, 2024ARA&A..62..529T}, the underlying driving microphysics~\citep{2020MNRAS.492.3179Y}, and environmental factors --- particularly the conditions of the surrounding halo.

Hot halo gas exerts inward pressure. This can oppose the development of a galactic outflow by reducing its velocity and limiting its extension compared to systems without a halo~\citep[e.g.][]{2021ApJ...917...12S}. By confining metal-enriched outflows and restricting the dispersal of ejecta, halo gas ram pressure has been considered to be instrumental in regulating baryonic recycling flows and enabling the enrichment of galaxies' CGM~\citep{2005ApJ...634L..37F}. 
In addition to this thermal pressure from the hot gas, galaxy halos may also host a reservoir of CRs. These CRs may originate as a relic population that could be transported by advection, bubbles associated with outflows, or the activity of a central supermassive black hole 
~\citep[see e.g.][]{2019MNRAS.484.1645O, 2021ApJ...914..135R, 2022ApJ...926....8S}.

Halo CRs can modify the structure of the circumgalactic medium (CGM)  and alter baryonic flows within galaxy halos~\citep[for reviews, see][]{2023A&ARv..31....4R, 2023Galax..11...86O}. Simulations of Milky-Way-mass galaxies suggest that CRs provide additional pressure support to sustain a multi-phase halo gas structure at low temperatures and can propel cool gas out to 100s of kpc~\citep{2018ApJ...868..108B, 2020MNRAS.496.4221J}. CR-driven winds can even push gas beyond the virial radius~\citep{2025OJAp....8E..66Q}. Due to the long CR survival time in halos, these feedback effects can continue to manifest long after the end of the mechanical processes that originally generated the CRs~\citep{2025OJAp....8E..66Q}. Halo CRs can also operate alongside hot halo gas to provide an inward non-thermal pressure that counteracts developing outflows. This is illustrated in Fig.~\ref{fig:schematic_flows}, which compares the suppressive effect of galaxy halos and the implications for baryonic recycling.

In this study, we assess the role of an extended CR halo in modifying the development of galactic winds driven by CR and thermal gas pressure, and derive the criteria for the breakout of an outflow from a galaxy with a CR halo. Our model focuses on the asymptotic, steady-state behavior of outflows driven by continuous central feedback, with particular emphasis on spatial scales extending from a few kiloparsecs above the galactic disk to $\sim$\, 100\,kpc into the halo. The relevant timescales correspond to $\sim$\,10\,Myr for shock breakout and $\sim 10-100$\,Myr for the wind to approach its terminal velocity, depending on the dominant driving mechanism. While our approach does not capture the time-dependent, bursty nature of star formation or CR transport in evolving magnetic fields, it provides a physically motivated framework for understanding the long-term impact of CR halos on outflow dynamics.

\section{Starburst-driven Outflows in Galaxy Halos}
\label{sec:sb_outflows_halo_model}

The collective feedback from a central galactic starburst can initiate a blastwave, which may develop into a sustained galactic wind if the starburst activity is continuous. 
Typically, about one SN explodes per $\sim$100 M$_{\odot}$ of star formation, with the exact rate dependent on the choice of the stellar initial mass function \citep[e.g.][]{1999ApJS..123....3L}.
We can therefore link the supernova event rate 
$\mathcal{R}_{\text{SN}} = \mathcal{R}_{-3} \, \text{kyr}^{-1}$ to the SFR of a galaxy by $\mathcal{R}_{\text{SF}} \sim 100 \, \text{M}_{\odot} \, \mathcal{R}_{\text{SN}}$. 
Observationally, star formation activity is typically quantified using the SFR surface density, $\Sigma_{\text{SFR}}$. This can be related to $\mathcal{R}_{\text{SF}}$ by considering that most star formation contributing to the blastwave occurs within a cylindrical region with a radius comparable to the disk scale height, $\sim H_{\text{s}}$, of a galaxy (defined by eq.~\ref{eq:scale_height}), i.e. $\Sigma_{\text{SFR}} =\mathcal{R}_{\text{SF}} \, / \left(\pi H_{\text{s}}^2 \right) = \Sigma_{\text{SFR, 0}} \, \text{M}_{\odot}\, \text{yr}^{-1} \, \text{kpc}^{-2}$. 
As we discuss in App. \ref{app:timescales}, the time it takes to establish an outflow ranges from several 10s of Myr to several 100s of Myrs \citep[see also e.g.][]{2023MNRAS.522.1843R, 2024MNRAS.52710897G, 2025MNRAS.540.1462R}, while individual starburst episodes only last about 10 Myr \citep{2023MNRAS.522.1843R}. Thus it is important to note that the SFR entering our model should be interpreted as an \textit{average} over many starburst episodes that collectively drive an outflow.

The injection rates of mass ($\dot{M}_{\text{SB}} = M_{\text{ej}} \mathcal{R}_{\text{SN}}$, for $M_{\text{ej}} = M_{\text{ej, 0}} \, \text{M}_{\odot}$ as the typical supernova ejecta mass) and energy ($\dot{E}_{\text{SB}} = E_{\text{SN}} \mathcal{R}_{\text{SN}}$, for $E_{\text{SN}} = 10^{51}\, E_{51} \, \text{erg}$ as the typical mechanical energy supplied by a supernova) supplied to an expanding blastwave from a galactic starburst can be linked to the SFR through the mass- and energy-loading factors $\eta_{\text{m}} \equiv \dot{M}_{\text{SB}} / \mathcal{R}_{\text{SF}} = 0.01 M_{\text{ej, 0}}$ and $\eta_{\text{e}} \equiv \varepsilon_{\text{w}} \dot{E}_{\text{SB}} / \left(E_{\text{SN}} \mathcal{R}_{\text{SN}}\right) = \varepsilon_{\text{w}}$, respectively, where $\varepsilon_{w}$ is a thermalization efficiency factor accounting for energy dissipation in the system \citep[see e.g.][]{2016MNRAS.455.1830T, 2017ApJ...834...25K, 2024ApJ...960..100S}. 
Similarly, we account for CR energy-dissipation with $\varepsilon_{\text{w, CR}}$ which for simplicity we here set to $\varepsilon_{\text{w, CR}} = \varepsilon_{\text{w}} = \eta_{\text{e}}$.
For convenience, we introduce the scaling parameters $\eta_{\text{m, -2}} =\eta_{\text{m}}/0.01$,  $\eta_{\text{e, -2}} = \eta_{\text{e}} / 0.01$ and $\varepsilon_{\text{w, -2}} = \varepsilon_{\text{w}} / 0.01$. The supply of CRs to the wind  
is parametrized by $f_{\text{CR}}$, representing the CR energy fraction at the galactic mid-plane. 
In our model, these parameters are treated as constants, remaining fixed throughout the evolution of the outflow, and the flow is considered to be driven by the combination of 
central thermal and kinetic energy-injection, and CR pressure gradients. 

\subsection{CR halos and their effects on outflows} 

Several observational studies have suggested the presence of extended CR reservoirs in galactic halos. These include a $\gamma$-ray halo around M31 reaching to 100s of kpc, which likely traces an interacting population of hadronic CRs~\citep{2021ApJ...914..135R}, $\gamma$-ray emission originating from halo clouds at kpc heights around the Milky Way~\citep{2015ApJ...807..161T}, and kpc-scale synchrotron emission from edge-on galaxies~\citep[e.g.][]{2018A&A...615A..98M, 2019A&A...632A..10M}. It has also been proposed that diffuse X-ray emission from the halos of Milky Way, M31, and lower-mass galaxies could originate from inverse Compton scattering, driven by a leptonic CR population~\citep{2025OJAp....8E..78H}. 

The formation of CR halos is a consequence of CR production during galaxy evolution. The long energy loss times of hadronic CRs in these environments (see Appendix~\ref{sec:halo_timescales}) ensure that most of the CR energy density supplied to a galaxy halo during its development can survive to the present day. Galaxies with significant historical stellar mass buildup are expected to host rich CR halos, even if their current star formation activity is low. Observations in $\gamma$-rays tentatively support this distinction, with CR halos primarily reported around massive late-type galaxies, while lower-mass galaxies show no indications of hosting such structures~\citep{2024arXiv241002066P}. 

To assess whether a CR halo can influence a developing outflow, the CR pressure contributions from both the halo and the outflow can be compared at a given altitude, $z$ (see Appendix~\ref{sec:model}; eqs. \ref{eq:CR_pressure_ext} and \ref{eq:CR_pressure_in}). For CRs to drive an outflow, the outward CR pressure must exceed the inward pressure from the halo CRs. When external and internal CR pressures become comparable, the driving effect of CR pressure gradients diminishes. 

As illustrated in Fig.~\ref{fig:schematic_flows}, the presence of a CR halo is expected to frustrate slow outflows if its scale height significantly exceeds the altitude where external and internal CR pressures are equal, $z_{\text{CR}}$. In units of the disk scale height, $H_{\rm s}$, this CR pressure equilibrium height is given by: 
\begin{equation}
    z_{\text{CR}} \sim 1.5 \, \varepsilon_{\text{w, -2}}^{1/2}\, E_{51}^{1/2} \, \mathcal{R}_{-3}^{1/2} \sigma_{1}^{-2} v_{\infty, 2}^{-1/2}\, H_{\text{s}} ~,
    \label{eq:zscale_cr}
\end{equation} 
where $\sigma=10 \, \sigma_{1} \, \text{km/s}$ is the gas velocity dispersion, 
and $v_{\infty, 2}$ is the rescaled terminal flow velocity defined as $v_{\infty} = 100\, v_{\infty, 2} \, \text{km/s}$ at large distances from the galactic plane. Equation~\ref{eq:zscale_cr} indicates that $z_{\text{CR}}$ is typically located at low altitudes for most galaxies. This suggests that CR-driven outflows are easily suppressed by the presence of an extended CR halo, if the halo has a scale-height $\gg H_{s}$. Galactic winds in systems with a well-developed CR halo are therefore expected to experience suppression, with their driving primarily dependent on thermal and kinetic energy-injection rather than CR pressure. 

\subsection{Outflow breakout criterion and terminal velocity} 

\begin{figure}
 \centering
 \includegraphics[width=0.9\linewidth, clip=true]{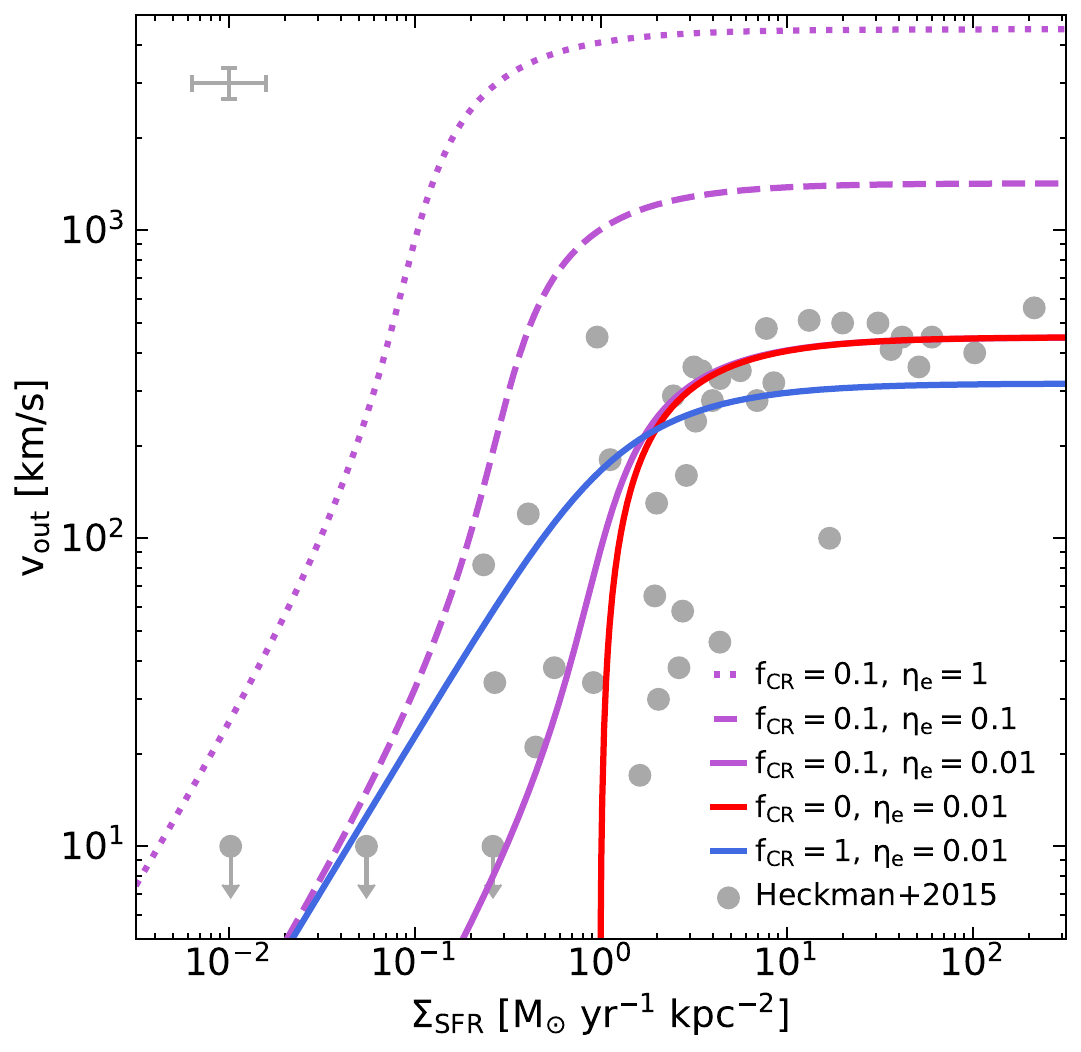}
 \caption{Terminal outflow velocities as a function of star formation rate surface density for different choices of CR energy fractions, $f_{\rm CR}$, and energy loading factors, $\eta_{\rm e}$. The model predictions are compared with data from \citet{2015ApJ...809..147H}, which show measured outflow velocities for a sample of nearby starburst galaxies with stellar masses in the range $\text{log}_{10} \left( M_{*} / \text{M}_{\odot}\right) \in [ 7.1 - 10.9 ]$. Typical uncertainties are indicated in the top-left corner. The model with $\eta_{\text{e}} \sim 0.01$ and $f_{\text{CR}} \sim 0.1$ provides a good match with the observed data. Models with higher energy loading values ($\eta_{\text{e}}$) generally predict outflow velocities in excess of the observations. The transition from slow outflows in weak starbursts to fast outflows in strong starbursts is best captured by models with $f_{\text{CR}} > 0$.
 } 
 \label{fig:CR_driven_outflows}
\end{figure}

For an outflow to break out from a galaxy, 
a minimum critical SFR surface density can be defined in the absence of a CR halo (see Appendix~\ref{app:outflow}). This is given by: 
\begin{equation}
     \Sigma_{\text{SFR, c}} \gtrsim 2.14 \, n_{0} \, \sigma_{1}^2 \, \eta_{\text{e, -2}}^{-1/2} \, \eta_{\text{m, -2}}^{-1/2}  ~~~\text{M}_{\odot}\, \text{yr}^{-1} \, \text{kpc}^{-2} ,  \label{eq:Sigma_crit}
\end{equation}
where $n_{\text{H, mp}} = n_{0} \, \text{cm}^{-3}$ is introduced as the mid-plane gas number density. 
In a strong star-starburst (i.e. when $\Sigma_{\text{SFR}} \gg \Sigma_{\text{SFR, c}}$), the maximum flow speed that can develop tends towards an asymptotic limit (see eq. \ref{eq:v_out}).  

CR-driven outflows can always be launched without a specific breakout criterion. However, in weak star formation scenarios (i.e. $\Sigma_{\text{SFR}} \ll \Sigma_{\text{SFR, c}}$), 
only very slow flow velocities can be achieved:  
\begin{equation}
     v_{\infty}^{\text{weak}} \rightarrow 237\, f_{\text{CR}} \eta_{\text{e, -2}}\, \Sigma_{\text{SFR, 0}} \, \sigma_{1}^{-2}\, \, n_{0}^{-1} ~~\text{km s}^{-1} \label{eq:v_out_weak} \ . 
\end{equation}
In this regime, the CR pressure equilibrium height reduces to: 
\begin{equation}
    z_{\text{CR}}^{\text{weak}} \rightarrow 1.85 f_{\text{CR}}^{-1/2} \, H_{\text{s}}\,
\end{equation}
which 
decreases as the CR supply to the system increases. This indicates that a CR halo strongly suppresses weak outflows that rely on CR driving (c.f. the right panel of Fig.~\ref{fig:schematic_flows}). 

In the strong starburst limit (when $\Sigma_{\text{SFR}} \gg \Sigma_{\text{SFR, c}}$), much faster terminal velocities are expected: 
\begin{equation}
    v_{\infty}^{\text{strong}} \rightarrow 10^{3} \, \eta_{\text{e, -2}}^{1/2} \, \eta_{\text{m, -2}}^{-1/2}\, \frac{\sqrt{1+f_{\text{CR}}} + \sqrt{1 - f_{\text{CR}}}}{2}  ~~\text{km s}^{-1}. \label{eq:v_out_strong} 
\end{equation}
The CR pressure equilibrium height in this regime is then  
\begin{equation}
    z_{\text{CR}}^{\text{strong}} \rightarrow 0.47 \, \varepsilon_{\text{w, -2}}^{1/4}\, E_{51}^{1/4} \, M_{\text{ej, 0}}^{1/4} \, \mathcal{R}_{-3}^{1/2} \sigma_{1}^{-2} \, H_{\text{s}}  \ . 
\end{equation} 
Although the inward halo CR pressure overtakes the outward flow-driving CR pressure near the disk scale height, 
the independence of $z_{\text{CR}}^{\text{strong}}$ from $f_{\text{CR}}$ suggests that outflows in this regime are momentum-dominated. 
Such outflows are unlikely to be significantly influenced by the presence of a CR halo, with thermal gas pressure likely playing a more critical role in regulating flow dynamics (see the central panel of Fig.~\ref{fig:schematic_flows}). 

The terminal velocities predicted by our model allow for comparison with observations. Figure~\ref{fig:CR_driven_outflows} shows terminal outflow velocities as a function of SFR surface density for a fiducial model with (dimensionless) parameters $n_{0} = 1$, $\sigma_{1} = 1$, $\eta_{\text{m, -2}} = 5$, and varying values of $f_{\text{CR}}$ and $\eta_\text{e, -2}$ as indicated in the legend. These calculations assume that all galaxies in the sample share a similar dynamical equilibrium pressure, $P_{\text{DE}} \sim \rho_{\text{mp}} \sigma^2 \sim G \Sigma^2$ \citep{2022ApJ...936..137O} for $\rho_{\text{mp}}$ as the galactic mid-plane gas volume density and $\Sigma$ as the corresponding surface density that sets $\Sigma_{\text{SFR, c}}$. For comparison, observed outflow velocities for a sample of nearby starburst galaxies with stellar masses in the range $\log_{10} (M_{*}/\text{M}_{\odot}) \in [7.1 - 10.9]$ are also shown~\citep{2015ApJ...809..147H}. 
Our model captures the general trend of observed flow velocities with parameter choices of $\varepsilon_{\text{w,-2}} \sim \eta_{\text{e,-2}} \sim 1$ and 
$\eta_{\text{m, -2}} \sim M_{\text{ej, 0}} \sim 5$. These choices align well with the energy and mass loading factors reported in numerical simulations of galactic outflows at altitudes of a few kpc \citep[e.g.][]{2018ApJ...853..173K, 2021MNRAS.504.1039R, 2024ApJ...960..100S, 2025OJAp....8E..66Q}. 

\section{Discussion and Implications}
\label{sec:discussion}

The structure of galactic winds has been extensively studied through theoretical approaches \citep[e.g.][]{1985Natur.317...44C, 2022ApJ...924...82F, 2023MNRAS.524.6374M}, 
detailed numerical simulations \citep[e.g.][]{2018ApJ...853..173K, 2023MNRAS.520.2655V, 2025OJAp....8E..66Q} and observational studies \citep[e.g.][]{2019ApJ...881...43K, 2023ApJ...956..142X, 2024ApJ...967...63B}. 
 These studies have shown that galactic winds are ubiquitous, particularly among star-forming galaxies, and that galaxies with high SFRs tend to drive faster and hotter winds. 
However, not all galaxies show clear signatures of outflows. Some systems show that gas launched from the ISM is recycled within a galaxy ecosystem rather than expelled \citep{2023A&A...670A..92M}. While the qualitative framework for wind launching is well established \citep[see][for a review]{2024ARA&A..62..529T}, few studies quantify outflow launching conditions \citep[e.g.][]{2015ApJ...809..147H, 2022ApJ...932...88O}. 

\citet{2015ApJ...809..147H} examined outflow speeds of galactic winds in a sample of nearby starbursts, and reported a sharp drop in flow velocities when the central SFR surface density fell below a critical threshold of $\Sigma_{\text{SFR, c}} \sim 1\,M_{\odot} \, \text{yr}^{-1} \, \text{kpc}^{-2}$. 
Our model shows that this critical SFR surface density arises due to the depth of the gravitational potential, which can only be overcome by sufficiently strong starbursts. The 
predicted value aligns with observations when considering typical dynamical equilibrium pressures set by the weight of the ISM, together with wind-loading parameters that are consistent with recent numerical studies of galactic winds in dwarf \citep{2024ApJ...960..100S} and spiral galaxies in the local Universe~\citep{2025OJAp....8E..66Q}.
However, we note that other studies find significantly higher mass and energy loading, depending on the methodology used \citep[e.g.][]{2015MNRAS.454.2691M, 2024MNRAS.527.1216S}.

\citet{2022ApJ...932...88O} investigated the feedback from star formation in a marginally Toomre-stable disk. They proposed that a young star cluster could launch an outflow if the starburst-driven shock reaches the disk scale height before its speed drops below the ISM's velocity dispersion. By assuming the formation of one star cluster per orbital timescale, they derived an expression for the critical SFR surface density, which is in rough agreement with the threshold found observationally by \citet{2015ApJ...809..147H}. However, their study does not account for the effects of gravity, which would alter the wind-launching criterion. 

Most numerical simulations of starburst-driven galactic winds show that the inclusion of CRs improves their ability to drive warm outflows with high mass-loading factors \citep[e.g.][]{2016ApJ...816L..19G, 2021MNRAS.504.1039R, 2022MNRAS.517..597C, 2024ApJ...964...99A}. This is in agreement with our results, which indicate that outflow speed in the CR-dominated regime is independent of the mass-loading factor, and suggest that CR-driven winds can sustain substantial mass-loading before being significantly slowed. However, current numerical models do not typically include a pre-existing CR halo in their initial conditions. Instead, they only model the accumulation of CRs within a galaxy ecosystem over time, usually as a consequence of stellar feedback. 
The widespread findings of substantial feedback impacts from CR-driven outflows in the literature may reflect this limitation --- a situation that has certain parallels with earlier numerical studies that lacked CGM thermal pressure, leading to unrealistically strong outflows in simulations of isolated galaxies \citep{2021ApJ...917...12S}.

We emphasize that CR-driven outflows are unlikely to operate efficiently in massive galaxies with deep gravitational potentials. Numerical studies \citep[e.g.][]{2018MNRAS.475..570J, 2024MNRAS.52710897G} show that in halos with $M_{\rm halo} \gtrsim 10^{12} \, M_\odot$, CR-driven winds are largely suppressed. This is due to a combination of strong gravitational binding, which raises the threshold for escape, and enhanced CR energy losses via Coulomb and hadronic interactions, which limit the dynamical role of CR pressure. These findings are consistent with our model, in which the outward CR pressure gradient becomes ineffective in the presence of an established CR halo, particularly when the halo’s scale height greatly exceeds the CR pressure equilibrium height. We thus interpret the CR suppression mechanism in our model as particularly relevant for lower-mass or early-phase starburst galaxies, while in more massive systems, thermal or radiation-driven mechanisms likely dominate the wind launching process.

At high redshift, several studies have reported high outflow velocities up to 1,000 $\text{km s}^{-1}$ \citep{2019ApJ...886...29S, 2022ApJ...933..222X}.
In general, higher outflow speeds require higher energy loading factors and lower mass loading factors. 
Observations suggest that high-redshift galaxies tend to be compact and turbulent \citep{2023ApJ...957...48G}, with high gas fractions and surface densities \citep[e.g.][]{2011ApJ...733..101G}, as well as low metallicities \citep{2008A&A...488..463M}.
Since the critical SFR surface density scales with the dynamical equilibrium pressure as $\dot{\Sigma}_{\text{SFR, c}}\propto P_{\text{DE}} \propto \Sigma^2$, thermally-driven outflows become ineffective in a highly turbulent, high surface-density environment. 
On the other hand, lower metallicities result in longer cooling times, leading to higher momentum loading \citep{2022ApJS..262....9O}. 
If mass-loading factors in high-redshift galaxies are comparable to those in low-redshift galaxies, then lower metallicities could explain the high observed outflow velocities in high-redshift galaxies, provided that the increased ISM weight does not suppress outflows. 

Curiously, the critical SFR surface density exhibits a stronger dependence on gas surface density than on the SFR surface density itself, which follows  $\dot{\Sigma}_{\text{SFR}} \propto \Sigma^{1.4}$ \citep{1989ApJ...344..685K}. This suggests that, counterintuitively, 
fast thermally-powered galactic outflows are expected to be more common at low surface densities, such as in dwarf galaxies, while being suppressed in extreme star-forming environments --- including massive clumps in high-redshift galaxies \citep{2011ApJ...733..101G} and the proposed feedback-free starburst galaxies at $z\sim 10$ \citep{2023ApJ...946L..13F, 2023MNRAS.523.3201D}. At high redshifts, CR-driven slow outflows may be more prevalent in these systems, as there would not have been sufficient time for them to establish a CR halo capable of suppressing outflows. Indeed, highly mass-loaded outflows are essential for regulating star formation and explaining the observed metallicity trends at high redshift \citep{2025MNRAS.tmp.1164T}, which can be naturally accounted for by CR-driven outflows.

We note that our model is intended as an idealized, steady-state framework that captures the asymptotic behavior of outflows under continuous feedback. While this does not resolve the detailed time-dependent evolution of bursty star formation or episodic feedback cycles, it provides a useful time-averaged description representative of either (i) long-duration starburst episodes or (ii) the cumulative effects of successive, shorter bursts over galactic dynamical timescales. A full treatment of the time dependence—including intermittent injection, CR streaming and diffusion, and adiabatic losses in 3D—will be essential in future work to extend the applicability of our model to more realistic galactic evolution scenarios.

\section{Conclusions}
\label{sec:conclusions}

In this study, we constructed a galactic outflow model driven by a continuous central feedback source, including the effects of CR pressure in the outflow and surrounding galaxy halo. We applied this model to a starburst galaxy to assess how the presence of a CR halo may influence outflow development. We found: 
\begin{enumerate}
  \item In the absence of CRs, galactic outflows are only launched if the SFR surface density exceeds a critical threshold proportional to the dynamic equilibrium pressure. At high SFR surface densities, these momentum-driven outflows approach the ejecta speed, reaching up to 1,000s of $\text{km s}^{-1}$.
  \item CRs can always drive slow outflows. We identified two different regimes: slow, CR-dominated outflows at SFR surface densities below the critical threshold, and fast, momentum-driven outflows at high SFR surface densities.
  \item In the presence of an extended CR halo, CRs become ineffective in sustaining outflows beyond the galactic scale height, leading to the suppression of CR-driven winds. 
\end{enumerate}
While our simplified approach is subject to substantial limitations (see Appendix~\ref{app:limitations}), it provides useful insights into the qualitative behavior of starburst-driven outflows and the influence of a CR halo. However, more detailed studies - including numerical simulations with CR halos as an initial condition - are needed to properly explore the physical impacts of CRs on the dynamical processes within galaxy halos. 

\begin{acknowledgements}
      We thank the anonymous referee for their insightful comments and suggestions that helped to improve the quality of this work.
      This research was funded in part by the Deutsche Forschungsgemeinschaft (DFG, German Research Foundation) under Germany's Excellence Strategy – EXC 2094 – 390783311. E.R.O is an international research fellow under the Postdoctoral Fellowship of the Japan Society for the Promotion of Science (JSPS), supported by JSPS KAKENHI Grant Number JP22F22327, and also acknowledges support from the RIKEN Special Postdoctoral Researcher Program for junior scientists. 
      This work was supported in part by the MEXT/JSPS KAKENHI grant numbers 20H00180, 22K21349, 24H00002, and 24H00241 (K.N.). K.N. acknowledges the support from the Kavli IPMU, the World Premier Research Centre Initiative (WPI), UTIAS, the University of Tokyo.  
\end{acknowledgements}

\bibliographystyle{aa} 
\bibliography{bibliography}

\begin{thebibliography}{92}
\expandafter\ifx\csname natexlab\endcsname\relax\def\natexlab#1{#1}\fi

\bibitem[{{Armillotta} {et~al.}(2024){Armillotta}, {Ostriker}, {Kim}, \& {Jiang}}]{2024ApJ...964...99A}
{Armillotta}, L., {Ostriker}, E.~C., {Kim}, C.-G., \& {Jiang}, Y.-F. 2024, \apj, 964, 99

\bibitem[{{Barcos-Mu{\~n}oz} {et~al.}(2018){Barcos-Mu{\~n}oz}, {Aalto}, {Thompson}, {Sakamoto}, {Mart{\'\i}n}, {Leroy}, {Privon}, {Evans}, \& {Kepley}}]{2018ApJ...853L..28B}
{Barcos-Mu{\~n}oz}, L., {Aalto}, S., {Thompson}, T.~A., {et~al.} 2018, \apjl, 853, L28

\bibitem[{{Behrendt} {et~al.}(2015){Behrendt}, {Burkert}, \& {Schartmann}}]{2015MNRAS.448.1007B}
{Behrendt}, M., {Burkert}, A., \& {Schartmann}, M. 2015, \mnras, 448, 1007

\bibitem[{{Bolatto} {et~al.}(2024){Bolatto}, {Levy}, {Tarantino}, {Boyer}, {Fisher}, {Cronin}, {Leroy}, {Klessen}, {Smith}, {Berg}, {B{\"o}ker}, {Boogaard}, {Ostriker}, {Thompson}, {Ott}, {Lenki{\'c}}, {Lopez}, {Dale}, {Veilleux}, {van der Werf}, {Glover}, {Sandstrom}, {Skillman}, {Chisholm}, {Villanueva}, {Lai}, {Lopez}, {Mills}, {Emig}, {Armus}, {Mayya}, {Meier}, {De Looze}, {Herrera-Camus}, {Walter}, {Rela{\~n}o}, {Koziol}, {Marvil}, {Jim{\'e}nez-Donaire}, \& {Martini}}]{2024ApJ...967...63B}
{Bolatto}, A.~D., {Levy}, R.~C., {Tarantino}, E., {et~al.} 2024, \apj, 967, 63

\bibitem[{{Bolatto} {et~al.}(2013){Bolatto}, {Warren}, {Leroy}, {Walter}, {Veilleux}, {Ostriker}, {Ott}, {Zwaan}, {Fisher}, {Weiss}, {Rosolowsky}, \& {Hodge}}]{2013Natur.499..450B}
{Bolatto}, A.~D., {Warren}, S.~R., {Leroy}, A.~K., {et~al.} 2013, \nat, 499, 450

\bibitem[{{Bradshaw} {et~al.}(2013){Bradshaw}, {Almaini}, {Hartley}, {Smith}, {Conselice}, {Dunlop}, {Simpson}, {Chuter}, {Cirasuolo}, {Foucaud}, {McLure}, {Mortlock}, \& {Pearce}}]{2013MNRAS.433..194B}
{Bradshaw}, E.~J., {Almaini}, O., {Hartley}, W.~G., {et~al.} 2013, \mnras, 433, 194

\bibitem[{{Buck} {et~al.}(2020){Buck}, {Pfrommer}, {Pakmor}, {Grand}, \& {Springel}}]{2020MNRAS.497.1712B}
{Buck}, T., {Pfrommer}, C., {Pakmor}, R., {Grand}, R. J.~J., \& {Springel}, V. 2020, \mnras, 497, 1712

\bibitem[{{Butsky} \& {Quinn}(2018)}]{2018ApJ...868..108B}
{Butsky}, I.~S. \& {Quinn}, T.~R. 2018, \apj, 868, 108

\bibitem[{{Cazzoli} {et~al.}(2016){Cazzoli}, {Arribas}, {Maiolino}, \& {Colina}}]{2016A&A...590A.125C}
{Cazzoli}, S., {Arribas}, S., {Maiolino}, R., \& {Colina}, L. 2016, \aap, 590, A125

\bibitem[{{Chan} {et~al.}(2022){Chan}, {Kere{\v{s}}}, {Gurvich}, {Hopkins}, {Trapp}, {Ji}, \& {Faucher-Gigu{\`e}re}}]{2022MNRAS.517..597C}
{Chan}, T.~K., {Kere{\v{s}}}, D., {Gurvich}, A.~B., {et~al.} 2022, \mnras, 517, 597

\bibitem[{{Chan} {et~al.}(2019){Chan}, {Kere{\v{s}}}, {Hopkins}, {Quataert}, {Su}, {Hayward}, \& {Faucher-Gigu{\`e}re}}]{Chan2019MNRAS}
{Chan}, T.~K., {Kere{\v{s}}}, D., {Hopkins}, P.~F., {et~al.} 2019, \mnras, 488, 3716

\bibitem[{{Chevalier} \& {Clegg}(1985)}]{1985Natur.317...44C}
{Chevalier}, R.~A. \& {Clegg}, A.~W. 1985, \nat, 317, 44

\bibitem[{{Cicone} {et~al.}(2016){Cicone}, {Maiolino}, \& {Marconi}}]{2016A&A...588A..41C}
{Cicone}, C., {Maiolino}, R., \& {Marconi}, A. 2016, \aap, 588, A41

\bibitem[{{Dekel} {et~al.}(2023){Dekel}, {Sarkar}, {Birnboim}, {Mandelker}, \& {Li}}]{2023MNRAS.523.3201D}
{Dekel}, A., {Sarkar}, K.~C., {Birnboim}, Y., {Mandelker}, N., \& {Li}, Z. 2023, \mnras, 523, 3201

\bibitem[{{Dermer} \& {Menon}(2009)}]{Dermer2009herb}
{Dermer}, C.~D. \& {Menon}, G. 2009, {High Energy Radiation from Black Holes: Gamma Rays, Cosmic Rays, and Neutrinos}

\bibitem[{{Devine} \& {Bally}(1999)}]{Devine1999ApJ}
{Devine}, D. \& {Bally}, J. 1999, \apj, 510, 197

\bibitem[{{Di Matteo} {et~al.}(2008){Di Matteo}, {Bournaud}, {Martig}, {Combes}, {Melchior}, \& {Semelin}}]{2008A&A...492...31D}
{Di Matteo}, P., {Bournaud}, F., {Martig}, M., {et~al.} 2008, \aap, 492, 31

\bibitem[{{Ferrara} {et~al.}(2005){Ferrara}, {Scannapieco}, \& {Bergeron}}]{2005ApJ...634L..37F}
{Ferrara}, A., {Scannapieco}, E., \& {Bergeron}, J. 2005, \apjl, 634, L37

\bibitem[{{Fielding} \& {Bryan}(2022)}]{2022ApJ...924...82F}
{Fielding}, D.~B. \& {Bryan}, G.~L. 2022, \apj, 924, 82

\bibitem[{{Finkelstein} {et~al.}(2023){Finkelstein}, {Bagley}, {Ferguson}, {Wilkins}, {Kartaltepe}, {Papovich}, {Yung}, {Arrabal Haro}, {Behroozi}, {Dickinson}, {Kocevski}, {Koekemoer}, {Larson}, {Le Bail}, {Morales}, {P{\'e}rez-Gonz{\'a}lez}, {Burgarella}, {Dav{\'e}}, {Hirschmann}, {Somerville}, {Wuyts}, {Bromm}, {Casey}, {Fontana}, {Fujimoto}, {Gardner}, {Giavalisco}, {Grazian}, {Grogin}, {Hathi}, {Hutchison}, {Jha}, {Jogee}, {Kewley}, {Kirkpatrick}, {Long}, {Lotz}, {Pentericci}, {Pierel}, {Pirzkal}, {Ravindranath}, {Ryan}, {Trump}, {Yang}, {Bhatawdekar}, {Bisigello}, {Buat}, {Calabr{\`o}}, {Castellano}, {Cleri}, {Cooper}, {Croton}, {Daddi}, {Dekel}, {Elbaz}, {Franco}, {Gawiser}, {Holwerda}, {Huertas-Company}, {Jaskot}, {Leung}, {Lucas}, {Mobasher}, {Pandya}, {Tacchella}, {Weiner}, \& {Zavala}}]{2023ApJ...946L..13F}
{Finkelstein}, S.~L., {Bagley}, M.~B., {Ferguson}, H.~C., {et~al.} 2023, \apjl, 946, L13

\bibitem[{{Genzel} {et~al.}(2023){Genzel}, {Jolly}, {Liu}, {Price}, {Lee}, {F{\"o}rster Schreiber}, {Tacconi}, {Herrera-Camus}, {Barfety}, {Burkert}, {Cao}, {Davies}, {Dekel}, {Lee}, {Lutz}, {Naab}, {Neri}, {Nestor Shachar}, {Pastras}, {Pulsoni}, {Renzini}, {Schuster}, {Shimizu}, {Stanley}, {Sternberg}, \& {{\"U}bler}}]{2023ApJ...957...48G}
{Genzel}, R., {Jolly}, J.~B., {Liu}, D., {et~al.} 2023, \apj, 957, 48

\bibitem[{{Genzel} {et~al.}(2011){Genzel}, {Newman}, {Jones}, {F{\"o}rster Schreiber}, {Shapiro}, {Genel}, {Lilly}, {Renzini}, {Tacconi}, {Bouch{\'e}}, {Burkert}, {Cresci}, {Buschkamp}, {Carollo}, {Ceverino}, {Davies}, {Dekel}, {Eisenhauer}, {Hicks}, {Kurk}, {Lutz}, {Mancini}, {Naab}, {Peng}, {Sternberg}, {Vergani}, \& {Zamorani}}]{2011ApJ...733..101G}
{Genzel}, R., {Newman}, S., {Jones}, T., {et~al.} 2011, \apj, 733, 101

\bibitem[{{Girichidis} {et~al.}(2016){Girichidis}, {Naab}, {Walch}, {Hanasz}, {Mac Low}, {Ostriker}, {Gatto}, {Peters}, {W{\"u}nsch}, {Glover}, {Klessen}, {Clark}, \& {Baczynski}}]{2016ApJ...816L..19G}
{Girichidis}, P., {Naab}, T., {Walch}, S., {et~al.} 2016, \apjl, 816, L19

\bibitem[{{Girichidis} {et~al.}(2024){Girichidis}, {Werhahn}, {Pfrommer}, {Pakmor}, \& {Springel}}]{2024MNRAS.52710897G}
{Girichidis}, P., {Werhahn}, M., {Pfrommer}, C., {Pakmor}, R., \& {Springel}, V. 2024, \mnras, 527, 10897

\bibitem[{{Heckman} {et~al.}(2015){Heckman}, {Alexandroff}, {Borthakur}, {Overzier}, \& {Leitherer}}]{2015ApJ...809..147H}
{Heckman}, T.~M., {Alexandroff}, R.~M., {Borthakur}, S., {Overzier}, R., \& {Leitherer}, C. 2015, \apj, 809, 147

\bibitem[{{Herenz} {et~al.}(2025){Herenz}, {Kusakabe}, \& {Maulick}}]{2025PASJ..tmp...80H}
{Herenz}, E.~C., {Kusakabe}, H., \& {Maulick}, S. 2025, \pasj [\eprint[arXiv]{2502.16969}]

\bibitem[{{Hopkins} {et~al.}(2023){Hopkins}, {Butsky}, {Ji}, \& {Kere{\v{s}}}}]{2023MNRAS.522.2936H}
{Hopkins}, P.~F., {Butsky}, I.~S., {Ji}, S., \& {Kere{\v{s}}}, D. 2023, \mnras, 522, 2936

\bibitem[{{Hopkins} {et~al.}(2022){Hopkins}, {Butsky}, {Panopoulou}, {Ji}, {Quataert}, {Faucher-Gigu{\`e}re}, \& {Kere{\v{s}}}}]{Hopkins2022MNRAS}
{Hopkins}, P.~F., {Butsky}, I.~S., {Panopoulou}, G.~V., {et~al.} 2022, \mnras, 516, 3470

\bibitem[{{Hopkins} {et~al.}(2025){Hopkins}, {Quataert}, {Ponnada}, \& {Silich}}]{2025OJAp....8E..78H}
{Hopkins}, P.~F., {Quataert}, E., {Ponnada}, S.~B., \& {Silich}, E. 2025, The Open Journal of Astrophysics, 8, 78

\bibitem[{{Irwin} {et~al.}(2024){Irwin}, {Beck}, {Cook}, {Dettmar}, {English}, {Heesen}, {Henriksen}, {Jiang}, {Li}, {Lu}, {Mele}, {M{\"u}ller}, {Murphy}, {Porter}, {Rand}, {Skeggs}, {Stein}, {Stein}, {Stil}, {Strong}, {Walterbos}, {Wang}, {Wiegert}, \& {Yang}}]{2024Galax..12...22I}
{Irwin}, J., {Beck}, R., {Cook}, T., {et~al.} 2024, Galaxies, 12, 22

\bibitem[{{Jacob} {et~al.}(2018){Jacob}, {Pakmor}, {Simpson}, {Springel}, \& {Pfrommer}}]{2018MNRAS.475..570J}
{Jacob}, S., {Pakmor}, R., {Simpson}, C.~M., {Springel}, V., \& {Pfrommer}, C. 2018, \mnras, 475, 570

\bibitem[{{Ji} {et~al.}(2020){Ji}, {Chan}, {Hummels}, {Hopkins}, {Stern}, {Kere{\v{s}}}, {Quataert}, {Faucher-Gigu{\`e}re}, \& {Murray}}]{2020MNRAS.496.4221J}
{Ji}, S., {Chan}, T.~K., {Hummels}, C.~B., {et~al.} 2020, \mnras, 496, 4221

\bibitem[{{Jones} {et~al.}(2019){Jones}, {Dowell}, {Lopez Rodriguez}, {Zweibel}, {Berthoud}, {Chuss}, {Goldsmith}, {Hamilton}, {Hanany}, {Harper}, {Lazarian}, {Looney}, {Michail}, {Morris}, {Novak}, {Santos}, {Sheth}, {Stacey}, {Staguhn}, {Stephens}, {Tassis}, {Trinh}, {Volpert}, {Werner}, {Wollack}, \& {HAWC+ Science Team}}]{2019ApJ...870L...9J}
{Jones}, T.~J., {Dowell}, C.~D., {Lopez Rodriguez}, E., {et~al.} 2019, \apjl, 870, L9

\bibitem[{{Kafexhiu} {et~al.}(2014){Kafexhiu}, {Aharonian}, {Taylor}, \& {Vila}}]{Kafexhiu2014PhRvD}
{Kafexhiu}, E., {Aharonian}, F., {Taylor}, A.~M., \& {Vila}, G.~S. 2014, \prd, 90, 123014

\bibitem[{{Karwin} {et~al.}(2019){Karwin}, {Murgia}, {Campbell}, \& {Moskalenko}}]{2019ApJ...880...95K}
{Karwin}, C.~M., {Murgia}, S., {Campbell}, S., \& {Moskalenko}, I.~V. 2019, \apj, 880, 95

\bibitem[{{Kennicutt}(1989)}]{1989ApJ...344..685K}
{Kennicutt}, Jr., R.~C. 1989, \apj, 344, 685

\bibitem[{{Kim} \& {Ostriker}(2015)}]{2015ApJ...802...99K}
{Kim}, C.-G. \& {Ostriker}, E.~C. 2015, \apj, 802, 99

\bibitem[{{Kim} \& {Ostriker}(2018)}]{2018ApJ...853..173K}
{Kim}, C.-G. \& {Ostriker}, E.~C. 2018, \apj, 853, 173

\bibitem[{{Kim} {et~al.}(2017){Kim}, {Ostriker}, \& {Raileanu}}]{2017ApJ...834...25K}
{Kim}, C.-G., {Ostriker}, E.~C., \& {Raileanu}, R. 2017, \apj, 834, 25

\bibitem[{{Koo} \& {McKee}(1990)}]{1990ApJ...354..513K}
{Koo}, B.-C. \& {McKee}, C.~F. 1990, \apj, 354, 513

\bibitem[{{Krieger} {et~al.}(2019){Krieger}, {Bolatto}, {Walter}, {Leroy}, {Zschaechner}, {Meier}, {Ott}, {Weiss}, {Mills}, {Levy}, {Veilleux}, \& {Gorski}}]{2019ApJ...881...43K}
{Krieger}, N., {Bolatto}, A.~D., {Walter}, F., {et~al.} 2019, \apj, 881, 43

\bibitem[{{Krumholz} {et~al.}(2020){Krumholz}, {Crocker}, {Xu}, {Lazarian}, {Rosevear}, \& {Bedwell-Wilson}}]{2020MNRAS.493.2817K}
{Krumholz}, M.~R., {Crocker}, R.~M., {Xu}, S., {et~al.} 2020, \mnras, 493, 2817

\bibitem[{{Lancaster} {et~al.}(2024){Lancaster}, {Ostriker}, {Kim}, {Kim}, \& {Bryan}}]{2024ApJ...970...18L}
{Lancaster}, L., {Ostriker}, E.~C., {Kim}, C.-G., {Kim}, J.-G., \& {Bryan}, G.~L. 2024, \apj, 970, 18

\bibitem[{{Laumbach} \& {Probstein}(1969)}]{1969JFM....35...53L}
{Laumbach}, D.~D. \& {Probstein}, R.~F. 1969, Journal of Fluid Mechanics, 35, 53

\bibitem[{{Lehnert} {et~al.}(1999){Lehnert}, {Heckman}, \& {Weaver}}]{1999ApJ...523..575L}
{Lehnert}, M.~D., {Heckman}, T.~M., \& {Weaver}, K.~A. 1999, \apj, 523, 575

\bibitem[{{Leitherer} {et~al.}(1999){Leitherer}, {Schaerer}, {Goldader}, {Delgado}, {Robert}, {Kune}, {de Mello}, {Devost}, \& {Heckman}}]{1999ApJS..123....3L}
{Leitherer}, C., {Schaerer}, D., {Goldader}, J.~D., {et~al.} 1999, \apjs, 123, 3

\bibitem[{{Leroy} {et~al.}(2015){Leroy}, {Walter}, {Martini}, {Roussel}, {Sandstrom}, {Ott}, {Weiss}, {Bolatto}, {Schuster}, \& {Dessauges-Zavadsky}}]{2015ApJ...814...83L}
{Leroy}, A.~K., {Walter}, F., {Martini}, P., {et~al.} 2015, \apj, 814, 83

\bibitem[{{Lopez-Rodriguez} {et~al.}(2021){Lopez-Rodriguez}, {Guerra}, {Asgari-Targhi}, \& {Schmelz}}]{2021ApJ...914...24L}
{Lopez-Rodriguez}, E., {Guerra}, J.~A., {Asgari-Targhi}, M., \& {Schmelz}, J.~T. 2021, \apj, 914, 24

\bibitem[{{Maiolino} {et~al.}(2008){Maiolino}, {Nagao}, {Grazian}, {Cocchia}, {Marconi}, {Mannucci}, {Cimatti}, {Pipino}, {Ballero}, {Calura}, {Chiappini}, {Fontana}, {Granato}, {Matteucci}, {Pastorini}, {Pentericci}, {Risaliti}, {Salvati}, \& {Silva}}]{2008A&A...488..463M}
{Maiolino}, R., {Nagao}, T., {Grazian}, A., {et~al.} 2008, \aap, 488, 463

\bibitem[{{Marasco} {et~al.}(2023){Marasco}, {Belfiore}, {Cresci}, {Lelli}, {Venturi}, {Hunt}, {Concas}, {Marconi}, {Mannucci}, {Mingozzi}, {McLeod}, {Kumari}, {Carniani}, {Vanzi}, \& {Ginolfi}}]{2023A&A...670A..92M}
{Marasco}, A., {Belfiore}, F., {Cresci}, G., {et~al.} 2023, \aap, 670, A92

\bibitem[{{Modak} {et~al.}(2023){Modak}, {Quataert}, {Jiang}, \& {Thompson}}]{2023MNRAS.524.6374M}
{Modak}, S., {Quataert}, E., {Jiang}, Y.-F., \& {Thompson}, T.~A. 2023, \mnras, 524, 6374

\bibitem[{{Mora-Partiarroyo} {et~al.}(2019){Mora-Partiarroyo}, {Krause}, {Basu}, {Beck}, {Wiegert}, {Irwin}, {Henriksen}, {Stein}, {Vargas}, {Heesen}, {Walterbos}, {Rand}, {Heald}, {Li}, {Kamieneski}, \& {English}}]{2019A&A...632A..10M}
{Mora-Partiarroyo}, S.~C., {Krause}, M., {Basu}, A., {et~al.} 2019, \aap, 632, A10

\bibitem[{{Mulcahy} {et~al.}(2018){Mulcahy}, {Horneffer}, {Beck}, {Krause}, {Schmidt}, {Basu}, {Chy{\.z}y}, {Dettmar}, {Haverkorn}, {Heald}, {Heesen}, {Horellou}, {Iacobelli}, {Nikiel-Wroczy{\'n}ski}, {Paladino}, {Scaife}, {Sridhar}, {Strom}, {Tabatabaei}, {Cantwell}, {Carey}, {Grainge}, {Hickish}, {Perrot}, {Razavi-Ghods}, {Scott}, \& {Titterington}}]{2018A&A...615A..98M}
{Mulcahy}, D.~D., {Horneffer}, A., {Beck}, R., {et~al.} 2018, \aap, 615, A98

\bibitem[{{Muratov} {et~al.}(2015){Muratov}, {Kere{\v{s}}}, {Faucher-Gigu{\`e}re}, {Hopkins}, {Quataert}, \& {Murray}}]{2015MNRAS.454.2691M}
{Muratov}, A.~L., {Kere{\v{s}}}, D., {Faucher-Gigu{\`e}re}, C.-A., {et~al.} 2015, \mnras, 454, 2691

\bibitem[{{Nianias} {et~al.}(2024){Nianias}, {Lim}, \& {Yeung}}]{2024ApJ...963...19N}
{Nianias}, J., {Lim}, J., \& {Yeung}, M. 2024, \apj, 963, 19

\bibitem[{{Oku} {et~al.}(2022){Oku}, {Tomida}, {Nagamine}, {Shimizu}, \& {Cen}}]{2022ApJS..262....9O}
{Oku}, Y., {Tomida}, K., {Nagamine}, K., {Shimizu}, I., \& {Cen}, R. 2022, \apjs, 262, 9

\bibitem[{{Orr} {et~al.}(2022){Orr}, {Fielding}, {Hayward}, \& {Burkhart}}]{2022ApJ...932...88O}
{Orr}, M.~E., {Fielding}, D.~B., {Hayward}, C.~C., \& {Burkhart}, B. 2022, \apj, 932, 88

\bibitem[{{Ostriker} \& {Kim}(2022)}]{2022ApJ...936..137O}
{Ostriker}, E.~C. \& {Kim}, C.-G. 2022, \apj, 936, 137

\bibitem[{{Ostriker} \& {McKee}(1988)}]{1988RvMP...60....1O}
{Ostriker}, J.~P. \& {McKee}, C.~F. 1988, Reviews of Modern Physics, 60, 1

\bibitem[{{Owen} {et~al.}(2018){Owen}, {Jacobsen}, {Wu}, \& {Surajbali}}]{Owen2018MNRAS}
{Owen}, E.~R., {Jacobsen}, I.~B., {Wu}, K., \& {Surajbali}, P. 2018, \mnras, 481, 666

\bibitem[{{Owen} {et~al.}(2019){Owen}, {Jin}, {Wu}, \& {Chan}}]{2019MNRAS.484.1645O}
{Owen}, E.~R., {Jin}, X., {Wu}, K., \& {Chan}, S. 2019, \mnras, 484, 1645

\bibitem[{{Owen} {et~al.}(2023){Owen}, {Wu}, {Inoue}, {Yang}, \& {Mitchell}}]{2023Galax..11...86O}
{Owen}, E.~R., {Wu}, K., {Inoue}, Y., {Yang}, H. Y.~K., \& {Mitchell}, A. M.~W. 2023, Galaxies, 11, 86

\bibitem[{{Pshirkov} \& {Nizamov}(2024)}]{2024arXiv241002066P}
{Pshirkov}, M.~S. \& {Nizamov}, B.~A. 2024, arXiv e-prints, arXiv:2410.02066

\bibitem[{{Quataert} \& {Hopkins}(2025)}]{2025OJAp....8E..66Q}
{Quataert}, E. \& {Hopkins}, P.~F. 2025, The Open Journal of Astrophysics, 8, 66

\bibitem[{{Rathjen} {et~al.}(2021){Rathjen}, {Naab}, {Girichidis}, {Walch}, {W{\"u}nsch}, {Dinnbier}, {Seifried}, {Klessen}, \& {Glover}}]{2021MNRAS.504.1039R}
{Rathjen}, T.-E., {Naab}, T., {Girichidis}, P., {et~al.} 2021, \mnras, 504, 1039

\bibitem[{{Rathjen} {et~al.}(2023){Rathjen}, {Naab}, {Walch}, {Seifried}, {Girichidis}, \& {W{\"u}nsch}}]{2023MNRAS.522.1843R}
{Rathjen}, T.-E., {Naab}, T., {Walch}, S., {et~al.} 2023, \mnras, 522, 1843

\bibitem[{{Rathjen} {et~al.}(2025){Rathjen}, {Walch}, {Naab}, {N{\"u}rnberger}, {W{\"u}nsch}, {Seifried}, \& {Glover}}]{2025MNRAS.540.1462R}
{Rathjen}, T.-E., {Walch}, S., {Naab}, T., {et~al.} 2025, \mnras, 540, 1462

\bibitem[{{Recchia} {et~al.}(2021){Recchia}, {Gabici}, {Aharonian}, \& {Niro}}]{2021ApJ...914..135R}
{Recchia}, S., {Gabici}, S., {Aharonian}, F.~A., \& {Niro}, V. 2021, \apj, 914, 135

\bibitem[{{Ruszkowski} \& {Pfrommer}(2023)}]{2023A&ARv..31....4R}
{Ruszkowski}, M. \& {Pfrommer}, C. 2023, \aapr, 31, 4

\bibitem[{{Shimoda} \& {Inutsuka}(2022)}]{2022ApJ...926....8S}
{Shimoda}, J. \& {Inutsuka}, S.-i. 2022, \apj, 926, 8

\bibitem[{{Shin} {et~al.}(2021){Shin}, {Kim}, \& {Oh}}]{2021ApJ...917...12S}
{Shin}, E.-J., {Kim}, J.-H., \& {Oh}, B.~K. 2021, \apj, 917, 12

\bibitem[{{Smith} {et~al.}(2024){Smith}, {Fielding}, {Bryan}, {Kim}, {Ostriker}, {Somerville}, {Stern}, {Su}, {Weinberger}, {Hu}, {Forbes}, {Hernquist}, {Burkhart}, \& {Li}}]{2024MNRAS.527.1216S}
{Smith}, M.~C., {Fielding}, D.~B., {Bryan}, G.~L., {et~al.} 2024, \mnras, 527, 1216

\bibitem[{{Steinwandel} {et~al.}(2024){Steinwandel}, {Kim}, {Bryan}, {Ostriker}, {Somerville}, \& {Fielding}}]{2024ApJ...960..100S}
{Steinwandel}, U.~P., {Kim}, C.-G., {Bryan}, G.~L., {et~al.} 2024, \apj, 960, 100

\bibitem[{{Su} {et~al.}(2010){Su}, {Slatyer}, \& {Finkbeiner}}]{Su2010ApJ}
{Su}, M., {Slatyer}, T.~R., \& {Finkbeiner}, D.~P. 2010, \apj, 724, 1044

\bibitem[{{Sugahara} {et~al.}(2019){Sugahara}, {Ouchi}, {Harikane}, {Bouch{\'e}}, {Mitchell}, \& {Blaizot}}]{2019ApJ...886...29S}
{Sugahara}, Y., {Ouchi}, M., {Harikane}, Y., {et~al.} 2019, \apj, 886, 29

\bibitem[{{Taylor} {et~al.}(2024){Taylor}, {Maltby}, {Almaini}, {Merrifield}, {Wild}, {Rowlands}, \& {Harrold}}]{2024MNRAS.535.1684T}
{Taylor}, E., {Maltby}, D., {Almaini}, O., {et~al.} 2024, \mnras, 535, 1684

\bibitem[{{Thompson} \& {Heckman}(2024)}]{2024ARA&A..62..529T}
{Thompson}, T.~A. \& {Heckman}, T.~M. 2024, \araa, 62, 529

\bibitem[{{Thompson} {et~al.}(2016){Thompson}, {Quataert}, {Zhang}, \& {Weinberg}}]{2016MNRAS.455.1830T}
{Thompson}, T.~A., {Quataert}, E., {Zhang}, D., \& {Weinberg}, D.~H. 2016, \mnras, 455, 1830

\bibitem[{{Tibaldo} {et~al.}(2015){Tibaldo}, {Digel}, {Casandjian}, {Franckowiak}, {Grenier}, {J{\'o}hannesson}, {Marshall}, {Moskalenko}, {Negro}, {Orlando}, {Porter}, {Reimer}, \& {Strong}}]{2015ApJ...807..161T}
{Tibaldo}, L., {Digel}, S.~W., {Casandjian}, J.~M., {et~al.} 2015, \apj, 807, 161

\bibitem[{{Toyouchi} {et~al.}(2025){Toyouchi}, {Yajima}, {Ferrara}, \& {Nagamine}}]{2025MNRAS.tmp.1164T}
{Toyouchi}, D., {Yajima}, H., {Ferrara}, A., \& {Nagamine}, K. 2025, \mnras [\eprint[arXiv]{2502.12538}]

\bibitem[{{Tsuru} {et~al.}(2007){Tsuru}, {Ozawa}, {Hyodo}, {Matsumoto}, {Koyama}, {Awaki}, {Fujimoto}, {Griffiths}, {Kilbourne}, {Matsushita}, {Mitsuda}, {Ptak}, {Ranalli}, \& {Yamasaki}}]{2007PASJ...59S.269T}
{Tsuru}, T.~G., {Ozawa}, M., {Hyodo}, Y., {et~al.} 2007, \pasj, 59, 269

\bibitem[{{Vasiliev} {et~al.}(2023){Vasiliev}, {Drozdov}, {Nath}, {Dettmar}, \& {Shchekinov}}]{2023MNRAS.520.2655V}
{Vasiliev}, E.~O., {Drozdov}, S.~A., {Nath}, B.~B., {Dettmar}, R.-J., \& {Shchekinov}, Y.~A. 2023, \mnras, 520, 2655

\bibitem[{{Veilleux} {et~al.}(2005){Veilleux}, {Cecil}, \& {Bland-Hawthorn}}]{2005ARA&A..43..769V}
{Veilleux}, S., {Cecil}, G., \& {Bland-Hawthorn}, J. 2005, \araa, 43, 769

\bibitem[{{Walter} {et~al.}(2017){Walter}, {Bolatto}, {Leroy}, {Veilleux}, {Warren}, {Hodge}, {Levy}, {Meier}, {Ostriker}, {Ott}, {Rosolowsky}, {Scoville}, {Weiss}, {Zschaechner}, \& {Zwaan}}]{2017ApJ...835..265W}
{Walter}, F., {Bolatto}, A.~D., {Leroy}, A.~K., {et~al.} 2017, \apj, 835, 265

\bibitem[{{Werhahn} {et~al.}(2023){Werhahn}, {Girichidis}, {Pfrommer}, \& {Whittingham}}]{2023MNRAS.525.4437W}
{Werhahn}, M., {Girichidis}, P., {Pfrommer}, C., \& {Whittingham}, J. 2023, \mnras, 525, 4437

\bibitem[{{Xu} {et~al.}(2022){Xu}, {Heckman}, {Henry}, {Berg}, {Chisholm}, {James}, {Martin}, {Stark}, {Aloisi}, {Amor{\'\i}n}, {Arellano-C{\'o}rdova}, {Bordoloi}, {Charlot}, {Chen}, {Hayes}, {Mingozzi}, {Sugahara}, {Kewley}, {Ouchi}, {Scarlata}, \& {Steidel}}]{2022ApJ...933..222X}
{Xu}, X., {Heckman}, T., {Henry}, A., {et~al.} 2022, \apj, 933, 222

\bibitem[{{Xu} {et~al.}(2023{\natexlab{a}}){Xu}, {Heckman}, {Henry}, {Berg}, {Chisholm}, {James}, {Martin}, {Stark}, {Hayes}, {Arellano-C{\'o}rdova}, {Carr}, {Huberty}, {Mingozzi}, {Scarlata}, \& {Sugahara}}]{2023ApJ...948...28X}
{Xu}, X., {Heckman}, T., {Henry}, A., {et~al.} 2023{\natexlab{a}}, \apj, 948, 28

\bibitem[{{Xu} {et~al.}(2023{\natexlab{b}}){Xu}, {Heckman}, {Yoshida}, {Henry}, \& {Ohyama}}]{2023ApJ...956..142X}
{Xu}, X., {Heckman}, T., {Yoshida}, M., {Henry}, A., \& {Ohyama}, Y. 2023{\natexlab{b}}, \apj, 956, 142

\bibitem[{{Yang} {et~al.}(2019){Yang}, {Gaspari}, \& {Marlow}}]{Yang2019ApJ}
{Yang}, H. Y.~K., {Gaspari}, M., \& {Marlow}, C. 2019, \apj, 871, 6

\bibitem[{{Yu} {et~al.}(2020){Yu}, {Owen}, {Wu}, \& {Ferreras}}]{2020MNRAS.492.3179Y}
{Yu}, B.~P.~B., {Owen}, E.~R., {Wu}, K., \& {Ferreras}, I. 2020, \mnras, 492, 3179

\bibitem[{{Zhang}(2018)}]{2018Galax...6..114Z}
{Zhang}, D. 2018, Galaxies, 6, 114

\bibitem[{{Zubovas} \& {Nayakshin}(2012)}]{Zubovas2012MNRAS}
{Zubovas}, K. \& {Nayakshin}, S. 2012, \mnras, 424, 666

\end{thebibliography}

\appendix

\section{Halo and Outflow model}
\label{sec:model}

\subsection{Halo model and CR timescales}
\label{sec:halo_timescales}

In our model, the galaxy halo consists of a thermal gas component and a non-thermal CR component. The gas is treated as an infinite slab in vertical hydrostatic equilibrium, with a density profile given by: 
\begin{equation}\label{eq:density_profile}
    \rho\left(z\right) = \rho_{\text{mp}} \, \text{cosh}^{-2}\left(\frac{z}{H_{\text{s}}}\right),
\end{equation}
and velocity dispersion 
$\sigma=10 \, \sigma_{1} \, \text{km s}^{-1}$, 
where $\rho_{\text{mp}} = \mu \, m_{\text{H}} \, n_{\text{H, mp}} $ is the mid-plane gas density, $\mu = 1.4$ is the mean atomic weight, $n_{\text{H, mp}} = n_{0} \, \text{cm}^{-3}$ is the number density of the gas, and where the disk scale-height is given by: 
\begin{equation}\label{eq:scale_height}
    H_{\text{s}} = \frac{\sigma}{\sqrt{2\pi G \rho_{\text{mp}}}}  \sim 338 \,\sigma_{1} n_{0}^{-0.5}\, \text{pc} 
\end{equation}
\citep{2015MNRAS.448.1007B}. The expected profile of the CR component in the halo is uncertain and depends on the underlying CR transport physics, which remain unsettled. 
It has been suggested that buoyant bubbles may redistribute CRs in the halo~\citep{2021ApJ...914..135R}. 
Such bubbles, blown by SN-powered winds in starburst regions of a galaxy~\citep[e.g.][]{2025PASJ..tmp...80H}, 
may be associated with outflows when they begin to fragment (e.g., above a cap similar to that seen in M82; \citealt{Devine1999ApJ}), or could even be attributed to buoyant structures inflated by intensive energetic outbursts akin to the Galactic \textit{Fermi} bubbles~\citep[e.g.][]{Su2010ApJ, Zubovas2012MNRAS, 2021ApJ...914..135R} analagous to AGN-inflated bubbles in Galaxy clusters~\citep[e.g.][]{Yang2019ApJ}. Observations of M31 do not yet strongly favor a particular CR distribution profile~\citep{2019ApJ...880...95K}, and detailed physical modeling of the CR halo is beyond the scope of the current work. 
We therefore adopt a simple spatially uniform CR distribution in the halo, effectively treating it as a CR bath in which an outflow develops. This approach is sufficient to obtain qualitative insights into the effects of a CR halo on outflow development, with more detailed modeling of CR transport mechanisms in galaxy halos left to future work. 

To achieve a sensible normalization at low altitudes where CR pressure most strongly influences outflow development, the CR content of the halo 
is set to match the CR energy fraction at the galactic mid-plane. The halo CR pressure can then be expressed as: 
\begin{equation} \label{eq:CR_pressure_ext}
    P_{\text{CR, ext}} = \frac{\gamma_{\text{CR}} - 1}{\gamma - 1} f_{\text{CR}} \rho_{\text{mp}} \sigma^2,
\end{equation}
where $f_{\text{CR}}$ is the fraction of energy supplied by the starburst to CRs at the galactic mid-plane, $\gamma = 5/3$ is the thermal gas adiabatic index, and $\gamma = 4/3$ is the CR fluid adiabatic index.
 This configuration creates a layered halo structure, with thermal gas pressure dominating at low altitudes and CR pressure becoming more important at higher altitudes (see the right panel of Fig.~\ref{fig:schematic_flows}). 

\begin{figure}
 \includegraphics[width=\linewidth, clip=true]{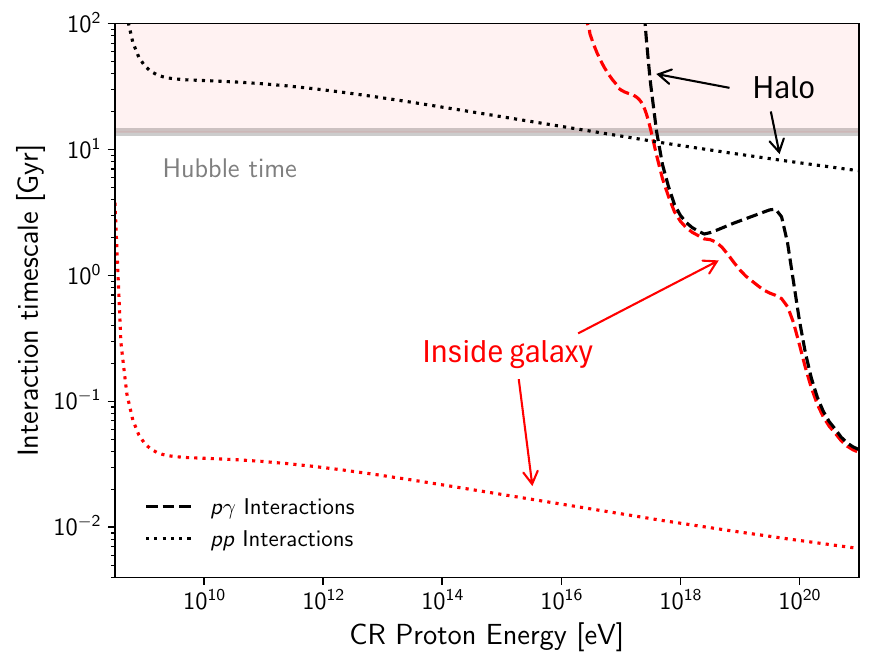}
 \caption{Characteristic timescales for a CR proton to undergo a hadronic interaction with ambient gas (pp interactions) or radiation (p$\gamma$ interactions) shown for typical conditions in the galaxy interior (red lines) and halo (black lines). For the galaxy interior, we adopt a gas density of $1 \, {\rm cm}^{-3}$ and stellar radiation fields consisting of a stellar component with $T = 7100$ K and an energy density of $\sim 0.7 \, {\rm eV \, cm}^{-3}$, with a dust component at $T = 60$ K and an energy density of $\sim 0.3 \, {\rm eV \, cm}^{-3}$. For halo conditions, we consider a reduced gas density of $10^{-3} \, {\rm cm}^{-3}$, with radiation energy densities scaled down by a factor of 100. p$\gamma$ losses with cosmological microwave background radiation at $z=0$ are included for both galaxy interior and halo conditions, with photo-pair and photo-pion interactions occurring at the same rate in both environments. The Hubble timescale is shown in gray. Interaction timescales exceeding this (i.e. within the shaded pink region) practically do not occur. Hadronic pp and p$\gamma$ interaction timescales are calculated following the approach of \citet{Owen2018MNRAS}, 
 assuming that the invariant energy of an interaction is dominated by that supplied by the CR. For pp interactions, this uses the total inelastic cross-section parametrization of~\cite{Kafexhiu2014PhRvD}. For p$\gamma$ pion-production losses, the step-function approximation to the cross-section proposed by~\cite{Dermer2009herb} is adopted. Bethe-Heitler losses are treated using the simplified cross-section from the same reference, but with fitting parameters updated according to~\cite{Owen2018MNRAS}, suitable for the energy range considered here.}
 \label{fig:CR_timescales}
\end{figure}

The CRs in the galaxy halo are likely primarily hadronic. This is because fast electron cooling in typical galactic conditions limits the electron population. Once deposited in the halo, CR hadrons experience few interaction or energy loss channels, allowing them to survive over Gyr timescales (see Fig.~\ref{fig:CR_timescales}). Adiabatic losses, streaming losses, and diffusive energy gains in micro-turbulence have only modest effects on the overall CR spectrum. Detailed simulations show that these processes rarely impact CR energies significantly in galaxy halos, typically contributing no more than 10 percent level corrections~\citep{Chan2019MNRAS, Hopkins2022MNRAS}. We therefore neglect these effects in our model. Without significant cooling or absorption channels, CR hadrons are expected to accumulate in the galaxy halo, forming a fossil record of the host galaxy's CR power generation history.

\subsection{Outflow model}\label{app:outflow}

We consider that an outflow is driven by a continuous injection of energy, momentum, CRs, and thermal gas, forming an expanding super-bubble that launches a galactic wind. This approach allows us to establish a criterion for wind launching and derive an analytical expression for the outflow velocity. 
To do this, we start from the blastwave equation of motion \citep{1988RvMP...60....1O} under the thin-shell and sector approximations \citep[e.g.][]{1969JFM....35...53L, 1990ApJ...354..513K}, to derive an analytically tractable equation of motion suitable for modeling CR-driven outflows:
\begin{equation}\label{eq:EoM}
    M \ddot{z} = \left(\Delta P + \Delta P_{\text{CR}} - \rho \dot{z}^2\right) z^2 + M\, g + \frac{\dot{M}_{\text{SB}}}{4\pi} \left( v_{\text{ej}} - \dot{z}\right) 
\end{equation}
Here $z = z\left(t\right)$ is the shock radius at time $t$ after the onset of shock expansion, $\Dot{z}$ and $\Ddot{z}$ are the instantaneous expansion velocity and acceleration of a thin shell following the expanding shock, 
$M = \int \rho z^2 dz$ is the swept-up mass, and $\Delta P$ and $\Delta P_{\text{CR}}$ are the thermal and non-thermal pressure differences between the shocked and the un-shocked gas, respectively. The gravitational acceleration, $ g < 0 $, for a single-component, isothermal slab in vertical hydrostatic equilibrium is given by: 
\begin{equation}
    g = - 2 \frac{\sigma^2}{H_{\text{s}}} \text{tanh}\left(\frac{z}{H_{\text{s}}}\right) \ . 
\end{equation}
The speed of the ejecta, $v_{\text{ej}}$, is: 
\begin{equation}
v_{\text{ej}} = \sqrt{2 \varepsilon_{w} \left(1 - f_{\text{CR}}\right) E_{\text{SN}} / M_{\text{ej}}} \ .
\end{equation}
Other terms retain the definitions provided in the main text (see section~\ref{sec:sb_outflows_halo_model}).

In the sector approximation \citep{1969JFM....35...53L}, we follow the one-zone dynamics of a shell-segment per unit angle, i.e. along a single streamline, which captures its dilution in three dimensions due to the growth of the surface area of the shell-segment as the shock expands.

We set $\Delta P = 0$ to account for the fact that the SBs generally become radiative, and quickly enter a rapidly-cooling wind phase \cite[e.g.][]{2015ApJ...802...99K, 2022ApJS..262....9O}.
While the expansion of such rapidly cooling winds is still formally energy-driven, the coupling between the hot interior and the shell leads to dynamics that are equivalent to those of a momentum-driven wind, but with a slightly boosted momentum injection rate in comparison to that at the source \citep{2024ApJ...970...18L}.
We account for this boost, as well as losses, by means of the energy efficiency factor $\varepsilon_{\text{w}}$.

We model CRs as a non-thermal fluid with an adiabatic index of $\gamma_{\text{CR}} = 4/3$, which is advected with the outflowing gas \citep{2024MNRAS.52710897G}, assuming a uniform pressure distribution immediately behind the shock. The CR pressure then evolves as: 
\begin{equation} \label{eq:CR_pressure_in}
    P_{\text{CR, in}} = 3 \left(\gamma_{\text{CR}} - 1\right) \frac{f_{\text{CR}}  \varepsilon_{\text{w, CR}}  \left(\dot{E}_{\text{SB}} / 4\pi\right) t}{z^3} \ ,
\end{equation} 
where $\varepsilon_{\text{w, CR}} = \zeta_{\rm str, CR} \;\! \zeta_{\rm cal, CR} \sim 0.01$ is an efficiency factor for the CR energy. 
This parametrizes the combined effects of energy losses due to 
CR streaming, $\zeta_{\rm str, CR} \sim 0.1$~\citep{Chan2019MNRAS, Hopkins2022MNRAS}, which accounts for their enhancement by stronger magnetic fields near the galactic disk \citep{2020MNRAS.497.1712B}, and CR attenuation by hadronic collisions, characterized by $\zeta_{\rm cal, CR}$.

Galactic starbursts are often CR calorimeters, meaning that a substantial amount of the CR energy is lost to hadronic pp collisions within their ISM~\citep{2020MNRAS.493.2817K, 2023MNRAS.525.4437W}. The CR calorimetric fraction has been reported to vary between 0.8 and 0.99 in nearby starbursts~\citep{2020MNRAS.493.2817K}. However, such strong calorimetry within the outflow itself would generally only arise before the outflow has cleared a path through the ISM of its host, i.e. the <10Myr break-out timescale (see Appendix~\ref{app:timescales}) compared with the >100 Myr duration of a 
wind-sustaining starburst~\citep[e.g.][]{2008A&A...492...31D}. A fiducial choice of $\zeta_{\rm cal, CR} \sim 0.1$ is therefore conservative. 

As the outflow reaches a steady state (i.e., over a timescale when the flow reaches its asymptotic limit at high altitudes), the mass it has swept up is given by:
\begin{align}
    M_{\infty} & = \int_0^{\infty} \rho\left(z\right) z^2 \text{d}z \\ \nonumber
    & = \frac{\pi^2}{12} \rho_{\text{mp}} H_{\text{s}}^3 \sim 10^{6} \, \sigma_{1}^{3} \, n_{0}^{-1/2} \, \text{M}_{\odot} \ . 
\end{align} 
To analyze the properties of the outflow, we consider steady-state solutions with $\Dot{z} \rightarrow v_{\infty} = 100\, v_{\infty, 2} \, \text{km s}^{-1}$ in the limit where $z = v_{\infty} t \gg H_{\text{s}}$, and where the mass of the shell is dominated by the swept-up mass, i.e., 
$\left(\Dot{M}_{\text{SB}} / 4\pi\right) t \ll M_{\infty}$.
In this limit, we can rewrite eq. \ref{eq:EoM} as: 
\begin{equation}\label{eq:EoM_asymptote}
    0 = \Delta P_{\text{CR}} z^2 - 2 M_{\infty} \frac{\sigma^2}{H_{\text{s}}} + \frac{\dot{M}_{\text{SB}}}{4\pi} \left(v_{\text{ej}} - v_{\infty}\right).
\end{equation} 
When external CR pressure is absent ($\Delta P_{\text{CR}} = P_{\text{CR, in}}$), this reduces to a quadratic form with a single positive solution: 
\begin{equation}\label{eq:v_out}
    v_{\infty} = \frac{v_{\text{ej}}}{2 \sqrt{1 - f_{\text{CR}}}} \left(\delta + \sqrt{\delta^2 + 2 \,  \left(\varepsilon_{\text{w, CR}} / \varepsilon_{\text{w}}\right) \, f_{\text{CR}}}\right),
\end{equation}
where 
\begin{equation}
    \delta = \sqrt{1 - f_{\text{CR}}} - \frac{\mathcal{R}_{\text{c}}}{\mathcal{R}_{\text{SN}}}
\end{equation}
and
\begin{equation}\label{eq:R_crit}
    \mathcal{R}_{\text{c}}^{-1} \sim 0.13 \, \varepsilon_{\text{w,-2}}^{1/2}\, E_{51}^{1/2} \, M_{\text{ej,0}}^{1/2} \, \sigma_{1}^{-4} \, \text{kyr} \ .
\end{equation}

In the absence of CRs, there are no outflow solutions when $\mathcal{R}_{\text{SN}} \leq \mathcal{R}_{\text{c}}$. On the other hand, when CRs are present, 
outflow solutions are always possible. For weak sources with $\mathcal{R}_{\text{SN}} \ll \mathcal{R}_{\text{c}}$, the outflow 
reaches very slow asymptotic speeds: 
\begin{equation}
    v_{\infty}^{\text{weak}} \rightarrow 66\, f_{\text{CR}}\,  \varepsilon_{\text{w, CR, -2}} \, E_{51}\, \mathcal{R}_{-3} \, \sigma_{1}^{-4}\, \text{km s}^{-1} \ , \label{eq:asymptotic_speed_weak}
\end{equation}
 where $\varepsilon_{w,{\rm CR},-2}\equiv \varepsilon_{w,{\rm CR}}/10^{-2}$ and 
${\mathcal R_{-3}}\equiv {\mathcal R_{\rm SN}}\,{\rm kyr}^{-1}.$

However, in the strong source limit where $\mathcal{R}_{\text{SN}} \gg \mathcal{R}_{\text{c}}$ and  $\delta\simeq\sqrt{1-f_{\rm CR}}$,  the outflow velocities are much higher, approaching: 
\begin{eqnarray}\label{eq:v_out_strong_micro}
    v_{\infty}^{\text{strong}} &\rightarrow&  5 \times 10^{2} \, \varepsilon_{w,-2}^{1/2} \, E_{51}^{1/2}\, M_{\text{ej, 0}}^{-1/2} \left( \sqrt{1 - f_{\text{CR}}} \right.\nonumber \\ 
    &&  +\left.\sqrt{1 + \left(2 \, \varepsilon_{\text{w, CR}} / \varepsilon_{\text{w}} - 1\right)f_{\text{CR}}}\right) \, \text{km s}^{-1} \ . 
\end{eqnarray}

\section{Timescales} \label{app:timescales}

To assess the validity of the underlying assumptions of our model, we check that the timescales related to the shock break-out from the ISM and the time it takes to reach the steady state solution are short enough to have a negligible effect on the overall dynamics.

The time it takes for a continuously driven SB to become radiative and dissipate the majority of its thermal energy is on the order of the shell-formation timescale \citep{2022ApJS..262....9O}
\begin{equation}
    t_{\text{sf}} \sim 0.12 \,\mathcal{R}_{-3}^{0.28}\,n_{0}^{-0.71}\,\text{Myr} ~,
\end{equation}
which is usually much shorter than the corresponding time it would take for an \textit{adiabatic} blastwave to reach the galactic scale height and break out; on the order of
\begin{equation}
    t_{\text{s}}^{\text{adiab.}} \sim 1.7\,\sigma_{1}^{5/3}\,n_{0}^{-1/2}\,\mathcal{R}_{\text{SN, }-3}^{-1/3}\,\text{Myr} ~.
\end{equation}
Thus, unless they are extremely powerful, the dynamics of starburst-driven SBs \textit{prior} to shock break-out can be described by a momentum-driven wind solution $z \propto t^{1/2}$ and the time it takes for the shock to break out is longer due to the radiative losses; on the order of
\begin{equation}
    t_{\text{s}}^{\text{rad.}} \sim 9.3\,\sigma_{1}^{2}\,n_{0}^{-1/2}\,\left(1-f_{\text{CR}}\right)^{-1/4}\, M_{\text{ej, }0}^{-1/4}\, \mathcal{R}_{\text{SN, }-3}^{-1/2}\,\text{Myr} ~.
\end{equation}
Prior to shock break-out the dynamics of the shock are well described by the expressions for momentum-driven winds in uniform media \citep{2022ApJS..262....9O, 2024ApJ...970...18L} in which the losses are explicitly accounted for. We can thus set $\varepsilon_{\text{w}} = 1$ for these pre-break-out expressions.
At the time of shock break-out the shock speed will have dropped dramatically to only about
\begin{equation}
    v_{\text{s}}^{\text{break-out}} \sim 18\, \sigma_{1}^{-1}\, \left(1-f_{\text{CR}}\right)^{1/4}\, M_{\text{ej, }0}^{1/4}\, \mathcal{R}_{-3}^{1/2}\,\text{km s}^{-1}~,
\end{equation}
and the shock is re-accelerated before it reaches its asymptotic speed. 
In the case of a weak CR-driven wind the acceleration after break-out is approximately $\propto v_{\text{s}}^{-1}$, leading to rapid acceleration up to the asymptotic speed Eq. \ref{eq:asymptotic_speed_weak} within only
\begin{equation}
    t_{\text{accel}}^{\text{weak}} \sim \frac{v_{\text{s}}}{\dot{v}_{\text{s}}} \sim 15 \, \sigma_{1}^{-1/2}\,n_{\text{0}}^{-1/2}\,\varepsilon_{\text{w, CR, }-2}^{-1/2}\,f_{\text{CR}}^{-1/2}\,M_{\text{ej, }0}^{3/8}\,\mathcal{R}_{-3}^{1/4}\,\text{Myr} ~,
\end{equation}
where we took the geometric average of the acceleration between $v_{\text{s}}^{\text{break-out}}$ and $v_{\infty}^{\text{weak}}$. In this time frame the outflow can already reach a few kpc above the disk.
In the strong-source limit the acceleration after break-out is $\propto \left(v_{\infty}^{\text{strong}} - v_{\text{s}}\right)$, which leads to a moderate, but still relatively fast acceleration to the asymptotic speed Eq. \ref{eq:v_out_strong_micro}, within
\begin{equation}
    t_{\text{accel}}^{\text{strong}} \sim 224\,\sigma_{1}^{2}\,n_{0}^{-1/2}\,\varepsilon_{\text{w, }-2}\,\left(1-f_{\text{CR}}\right)^{-1/4}\, M_{\text{ej, }0}^{-1/4}\, \mathcal{R}_{\text{SN, }-3}^{-1/2}\,\text{Myr} ~.
\end{equation}
In this time frame the wind may have already expanded many 10s of kpcs into the halo, while stronger winds may reach their asymptotic speed closer to the galaxy.

These timescales may be compared to the typical duration of a starburst episode on the order of $t_{\text{starburst}} \sim 10\,\text{Myr\,}$\citep{2023MNRAS.522.1843R}, which suggests that starburst driven winds are established over the course of \textit{multiple} short bursts, each lasting only a fraction of the lifetime of the wind. The SN rate $\mathcal{R_{\text{SN}}}$ in our model thus represents the \textit{average} over a series of distinct starbursts.

\section{Model Limitations and Assumptions} \label{app:limitations}

Our model invokes a number of approximations and assumptions. While we consider our results to be qualitatively robust, future developments that relax these assumptions may provide more refined insights. Here, we assess the validity of our assumptions and approximations and discuss their potential impact on our results.

\paragraph{Density Profile and Gravitational Field.} The density profile in eq.\,(\ref{eq:density_profile}) represents an isothermal, single-component atmosphere in vertical hydrostatic equilibrium, providing a reasonably accurate description near the mid-plane. However, the presence of molecular gas and a young stellar disk could create a deeper potential well, leading to a more compact density profile that may affect the early stages of shock breakout. Moreover, explicitly modeling the gravitational potential of different galaxy components (e.g. the bulge, disk and dark-matter halo) could modify flow properties \citep[see][]{2022ApJ...926....8S}, and this can be affected by the detailed gas distribution throughout the galaxy and inner halo. However, the outflow properties at high altitudes (above $\sim 100$ kpc) are unlikely to be significantly impacted if the overall gas surface density and wind loading remain unchanged. 

\paragraph{Halo Gas Properties.} In the halo, the multi-phase CGM contributes a diffuse gas background that exerts additional (ram) pressure, which can counteract outflow expansion \citep{2021ApJ...917...12S}.
While the CGM is diffuse, it is expected to follow a shallow density profile that would slow the outflow's expansion once the swept-up CGM mass becomes comparable to $M_{\infty}$. This occurs at a height of $z_{\text{CGM}} \sim 5 \, \sigma_{1} \, n_{0}^{-1/6} \, n_{\text{CGM, -3}}^{-1/3} \, \text{kpc}$, 
where $n_{\text{H, CGM}} = 10^{-3}\,n_{\text{CGM, -3}} \, \text{cm}^{-3}$ is the number density of hydrogen in the CGM.

\paragraph{Ejecta Mass.} For the asymptotic steady-state solution, we consider a limit where the outflow has traveled sufficiently far above the disk, yet is not old enough for its mass to be dominated by the ejecta. 
The timescale for the outflow to become ejecta-dominated is $  t_{\text{ED}} \sim 4\pi M_{\infty} / \Dot{M}_{\text{SB}} \sim 12.6 \, \sigma_{1}^3 \, n_{0}^{-1/2} \, M_{\text{ej, 0}}^{-1} \, \mathcal{R}_{-3}^{-1} \, \text{Gyr}$, which 
is considerably longer than the lifetime of the starburst driving the outflow. We therefore consider our results to be robust against this approximation. 

\paragraph{Thin-Shell and Sector Approximation.} While the thin-shell approximation is well-suited for radiative blastwaves or those propagating through media with positive density gradients, it becomes increasingly crude in environments with steep negative density gradients. This is because the mass, energy, and momentum distributions behind the shock broaden significantly \citep[see e.g.][]{1969JFM....35...53L, 1990ApJ...354..513K}. 
Despite the steep density gradient considered in this work, our model focuses on a radiative, momentum-driven wind, where the thin-shell approximation is expected to remain reasonably accurate. 
The sector approximation, on the other hand, is most reliable when the local shock surface remains relatively flat. This approximation can break down if adjacent streamlines begin to diverge significantly, e.g. due to sudden deflections.  However, since observed outflows generally exhibit large opening angles, this approximation is likely to have only a negligible impact on our findings. 

\paragraph{Treatment of CRs.} Our treatment of CRs involves a number of simplifications. While we do not explicitly account for CR cooling or interactions within the flow, this is well justified given the relevant timescales (see Appendix~\ref{sec:halo_timescales}), at least in the context of low mass galaxies. 
In more massive galaxies with denser gas, it becomes increasingly difficult for shocks driven by stellar feedback to break-out of the disk, where cooling losses do dominate, rendering CR-driven galactic outflows ineffective in such systems with halo masses above $\gtrsim 10^{12}\,\text{M}_{\odot}$ \citep{2018MNRAS.475..570J, 2024MNRAS.52710897G}.
However, future studies with more detailed CR propagation modeling may yield different quantitative results, particularly if a detailed CR transport model within the halo is included. Such models could alter the distribution of halo CRs~\citep{2021ApJ...914..135R}, 
and reveal the microphysical impacts of CRs on outflows, including re-acceleration processes and interactions with complex magnetic field structures at the outflow–halo interface~\citep[see, e.g.][]{2023MNRAS.522.2936H}. Additionally, our study does not consider the spectral evolution of CR particles, which could influence the coupling between CRs and the outflowing wind fluid, potentially modifying their driving efficiency and altering observable CR emission signatures. Future work incorporating these effects could provide a clearer understanding of how CRs shape the development of outflows in CR-rich galaxy halos.

\end{document}